\input psfig
\documentstyle[12pt,epsf]{article}
\begin{document}
\setlength{\baselineskip}{0.33 in}
\catcode`@=11
\long\def\@caption#1[#2]#3{\par\addcontentsline{\csname
  ext@#1\endcsname}{#1}{\protect\numberline{\csname
  the#1\endcsname}{\ignorespaces #2}}\begingroup
    \small
    \@parboxrestore
    \@makecaption{\csname fnum@#1\endcsname}{\ignorespaces #3}\par
  \endgroup}
\catcode`@=12
\newcommand{\newc}{\newcommand}
\newc{\gsim}{\lower.7ex\hbox{$\;\stackrel{\textstyle>}{\sim}\;$}}
\newc{\lsim}{\lower.7ex\hbox{$\;\stackrel{\textstyle<}{\sim}\;$}}
\newenvironment{fnitemize}{\begin{itemize}
	\footnotesize}{\end{itemize}}

\vspace*{-1.25in}
\vspace*{.85in}
\begin{center}
{\Large{\bf  Quenched Approximation Artifacts: A study in 2-dimensional QED}
}\\
\vspace*{.45in}
{\large{W.~Bardeen$^1$,A.~Duncan$^2$, E.~Eichten$^1$, and H.~Thacker$^3$}} \\ 
\vspace*{.15in}
$^1$Fermilab, P.O. Box 500, Batavia, IL 60510 \\
$^2$Dept. of Physics and Astronomy, University of Pittsburgh, Pittsburgh, PA 15620\\
$^3$Dept.of Physics, University of Virginia, Charlottesville, VA 22901
\end{center}
\vspace*{.3in}
\begin{abstract}
 The spectral properties of the Wilson-Dirac operator in 2-dimensional QED
 responsible for the appearance of exceptional configurations in quenched
 simulations are studied in detail.  The mass singularity structure of the
 quenched functional integral is shown to be extremely compicated, with multiple
 branch points and cuts. The connection of lattice topological charge and 
 exactly real eigenmodes is explored using cooling techniques.  The lattice
 volume and spacing dependence of these modes is studied, as is the effect
 of clover improvement of the action. A recently proposed modified quenched
 approximation is applied to the study of meson correlators, and the results
 compared with both naive quenched and full dynamical calculations of the
 same quantity.
\end{abstract}

\newpage

\section{Introduction}

   A basic quantity in essentially all lattice calculations of hadronic 
 properties is the quark propagator in the presence of a fixed external
 gauge field. Hadron propagation amplitudes or current matrix elements
 are extracted from averages of products of quark propagators over 
 ensembles of gauge fields generated relative to either a pure gluonic
 weight $e^{{\rm (pure\; gauge\; action)}}$ for the quenched approximation,
 or $e^{{\rm (pure\; gauge\; action)}}\cdot{\rm (determinant\; of\;Dirac\; operator)}$
 for a fully dynamical calculation including internal quark loop effects.
 There has naturally been considerable attention devoted to the issue of
 the extent to which the quenched approximation deviates from the full
 theory  \cite{qnchgen}. Are the differences of a detailed quantitative character only,
 perhaps mainly absorbable in a redefinition of the scale of the theory,
 or are there serious qualitative differences which make it difficult to
 extract reliable physics from a quenched calculation?

   Qualitative differences which distinguish the quenched theory 
from the full dynamics concern the sensitive dependence of certain 
amplitudes on the quark mass $m$ in
the chiral limit $m\rightarrow 0$.   Generic effects include the ``hairpin" 
diagrams which signal the generation of the singlet $\eta^{\prime}$ mass and the 
corresponding ``quenched chiral logs" \cite{chirlog} which are expected
to give logarithmic corrections to the mass dependence in the chiral limit.
In lattice calculations, finite volume effects and convergence problems 
in evaluating the light quark propagators are also elements which impact the
study of the chiral limit.   A particular artifact  
associated with quenched calculations involving 
light Wilson fermions is the occasional appearance of
``exceptional" gauge configurations \cite{exceptconf} with apparently 
nonsensical values for the light quark propagators resulting
in extremely noisy statistical averages and
unreliable values for the hadronic amplitudes.  Clearly, it will be 
impossible to determine, in any definitive way, the chiral structure of the
quenched theory until the nature of the
exceptional configuration problem 
 is fully understood and the resulting
pathologies tamed.  Our objective here is
to study, in a specific model - 2 dimensional QED-  under very precise 
numerical control, the role of the exceptional configuration phenomenon 
in quenched calculations with light Wilson fermions.  In the final section, 
we shall use this model to perform a series of numerical tests on
a  recently proposed cure of the exceptional configuration problem
\cite{qzpaper} which suggests that greatly improved results are indeed
possible for calculations using Wilson-Dirac fermions
in the low mass chiral regime.

     Following the example of Smit and Vink \cite{smitvink}, we can take advantage 
  of the fact that 
  2-dimensional QED (massive Schwinger model) provides a particularly useful testbed
  for investigating the pathologies of  the quenched approximation. As in
  4-dimensional QCD, there is an intimate connection between the topological
   structure of the theory and the spectrum of the Dirac operator near the origin,
  which controls the chiral limit behavior.  Moreover, proposed
  modifications of the quenched approximation can be tested against the full
  theory ({\em including} determinant effects) in great detail in the 2-dimensional case.

    On large lattices at  weak coupling the Wilson spectrum $\{\lambda_a\}$  settles (taking QED2
  as an example) near the real axis  into four distinct branches at ${\rm Re}(\lambda)=\pm 2, 0$, with the central
  region containing two branches. It is usual to select the undoubled left branch (see Fig(1))
 as a model
  for the continuum theory by tuning the bare quark parameter $m$ appropriately.  For
 example, in QED2 one finds \cite{smitvink} a critical line at  ${\rm Re}(\lambda)\approx -2+0.65/\beta$,
 with a dispersion of  the real part of the eigenvalue around this value which grows
 rapidly as one goes to stronger coupling.  In particular, at stronger coupling one finds
 an increasing density of exactly real eigenmodes straying to the left away from the critical line
 into the region corresponding to small positive quark masses. Consequently, evaluation
 of the propagator in configurations giving rise to such modes necessarily results in
  noisy results, as one occasionally ends up sitting very close to a pole of the propagator. In a sense,
 the original  rationale for moving from Minkowski to Euclidean space - avoiding the poles of the
 propagator at Minkowski momenta- has been vitiated by the complex (i.e. non-skewhermitian)
 character of the Wilson-Dirac operator.  The exceptional configurations observed
 in quenched calculations  appear to be due to precisely this feature of the Wilson
 action, and the study of these configurations reduces essentially to a study of the 
 frequency and location of the exactly
 real eigenvalues of the Wilson-Dirac matrix defining the quadratic fermion action of 
 a lattice theory.  

    The present paper is organized as follows. In Section 2 we review some of the basic
 properties of the Wilson-Dirac matrix in lattice gauge theory. Most of the results listed here
 are well known but the review allows us to clarify our notation and to remind the reader
  of some exact properties of the Wilson-Dirac spectrum following purely from linear-algebraic
 considerations. Results are given here both for QED in 2 and QCD in 
  4 dimensions. Section 3 contains a more detailed analysis of the mass singularity structure
 of the quenched functional integral. It is shown that the quenched functional integral
 as defined in a Monte Carlo simulation is strictly speaking undefined. The functional integral
 must be defined by analytic continuation from a pole-free region of the complex quark mass
 plane. In the region of physical interest, the path integral defines a cut function in this plane
 with a complicated branch structure.  Section 4 describes the relationship between topological
 charge and the exactly real modes responsible for the pathologies of the quenched 
 approximation.  In Section 5 we study the dependence of the real part of the Wilson-Dirac
 spectrum on lattice volume and lattice spacing.  In Section 6 we give the results of a detailed study of
 the changes in  the real spectrum  induced by the addition of a clover improvement term 
 \cite{clover} to
 the action. Finally, in Section 7 we test the validity of a recently proposed \cite{qzpaper}
modified
 quenched approximation (MQA) which resolves the singularities introduced by real
 eigenmodes by a simple pole-shifting procedure. In particular, we carry out an explicit comparison of the MQA with both the
 naive quenched and full dynamical results for correlators of quark bilinears.

\section{Formal properties of the Wilson-Dirac operator}
     The Euclidean Dirac operator  $D\!\!\!/(A)$ in the continuum is formally skew-hermitian
 for smooth gauge fields $A$, so at a formal level the quark propagator
 $(D\!\!\!/(A)+m)^{-1}$ exists for all nonzero quark mass $m$. The existence of
 the inverse at zero quark mass is related to the topological  structure of the 
 background gauge field by the Atiyah-Singer index theorem. For example, 
 in 2D QED, defining the Dirac operator on a compact  2-sphere of radius $R$ by the usual
 one-point compactification \cite{NielSchr} gives a rigorously self-adjoint
 $iD\!\!\!/(A)$ (for  gauge fields which approach pure gauge at Euclidean  infinity)
 with a purely discrete spectrum :
\begin{eqnarray}
 iD\!\!\!/(A)\chi_{a}(x)&=&\lambda_{a}\chi_{a}(x) \\
 \int d^{2}x\frac{2R}{x^2+R^2}\chi^{\dagger}_{a}(x)\chi_{b}(x)&=&\delta_{ab} \\
 \frac{e}{4\pi}\epsilon_{\mu\nu}F_{\mu\nu}+\frac{2R}{x^2+R^2}\sum_{a}\chi^{\dagger}_{a}
\gamma_{5}\chi_{a}(x)&=&{\rm pure\;divergence}
\end{eqnarray}

  The eigenfunctions $\chi_{i}$ are normalizable relative to the inner product (2), and
 the  ``pure divergence" referred to in (3) is  just the divergence of the axial vector
  current of the model, which is seen to reduce to a sum of a topological term and
  a sum over the $\gamma_5$ eigenvalues of each of the discrete modes. The latter
 sum (as a consequence of $\{\gamma_5,D\!\!\!/(A)\}=0$) reduces to a sum over 
 zero modes only, and therefore to a difference of the number of positive and negative chirality
 zero mode solutions of the Dirac equation in the background field. Nielsen and 
 Schroer have given an explicit analytic expression for these solutions in an 
 arbitrary smooth (and asymptotically pure-gauge) field.  Essentially the same statements
 hold for QCD in 4 dimensions, with the obvious replacement of $\epsilon_{\mu\nu}F_{\mu\nu}$
 by the appropriate 4-dimensional Chern-Simons $F\tilde{F}$ term. 

    Once the Euclidean space-time is discretized as in lattice theory, little of the
 preceding formalism remains in exact form, though qualitative remnants of the
 zero mode structure can be observed provided one works on large lattices with
 very smooth background gauge fields.   In the first place, although a discrete
 spectrum is now guaranteed (the Dirac operator is a finite matrix!), the addition of
 a Wilson term to eliminate doubling 
destroys the skew-hermitian nature of the Dirac operator. Specifically, quark 
 propagators are inverses of a matrix $D-rW+m\equiv {\cal M}+m$, with $D$, $W$ and $m$ the
 naive Dirac matrix, $W$ the Wilson term, and $m$ a quark mass parameter:

\begin{eqnarray}
\label{eq:Mdef}
 {\cal M}&\equiv& D-rW \\
\label{eq:Ddef}
 D_{a\alpha\vec{m},b\beta\vec{n}}&=&\frac{1}{2}(\gamma_{\mu})_{ab}U_{\alpha\beta}(\vec{m}\mu)\delta_{\vec{n},\vec{m}+\hat{\mu}}-\frac{1}{2}(\gamma_{\mu})_{ab}U^{\dagger}_{\alpha\beta}
(\vec{n}\mu)\delta_{\vec{n},\vec{m}-\hat{\mu}} \\
\label{eq:Wdef}
 W_{a\alpha\vec{m},b\beta\vec{n}}&=&\frac{1}{2}\delta_{ab}(U_{\alpha\beta}(\vec{m}\mu)\delta_{\vec{n},\vec{m}+\hat{\mu}}+U^{\dagger}_{\alpha\beta}(\vec{n}\mu)\delta_{\vec{n},\vec{m}-\hat{\mu}} )
\end{eqnarray}
Here $a,b$ are Dirac indices, $\alpha,\beta$ color indices, $\vec{m},\vec{n}$ lattice sites,
 and $r$ the Wilson parameter, which must be chosen $0<r\leq 1$ and is usually
 taken to be unity.  As $D$ is skew-hermitian and $W$ hermitian, the full Wilson-Dirac
${\cal M}= D-rW$ 
 matrix is complex, and the generic eigenvalue (which locates  poles of the
 quark propagator in the complex mass plane) will also be complex. 

 It will frequently be convenient to use a single index $i\equiv(a\alpha\vec{m})$
 to identify the single Grassmann variable for a quark field component, and to write
 the Wilson-Dirac matrix elements with the notation ${\cal M}_{ij}$.  From (\ref{eq:Wdef})
 it follows immediately that although ${\cal M}$ is neither skewhermitian (unless $r$=0)
 nor hermitian, the matrix $\gamma_5 {\cal M}$ is hermitian. Consequently the coefficients
 of the secular equation for ${\cal M}$ are real:
\begin{eqnarray}
 {\rm det}(\lambda - \gamma_5 {\cal M}) &=& ({\rm det}(\lambda^{*}-\gamma_5 {\cal M}))^{*} \\
  {\rm det}(\lambda - {\cal M}) &=& ({\rm det}(\lambda^{*}-{\cal M}))^{*}
\end{eqnarray}
so that if $\lambda$ is an eigenvalue of ${\cal M}$, so is $\lambda^{*}$- the eigenvalues
 are therefore either real or appear in conjugate pairs. The nearest neighbour
 structure of ${\cal M}$ also implies immediately that
\begin{equation}
\label{eq:noodd}
   {\rm Tr}({\cal M}^n)=0,\;\;\;n\;{\rm odd}
\end{equation}
which implies that odd powers of $\lambda$ are absent in the secular equation ${\rm det}(
\lambda-{\cal M})=0$. Consequently, if $\lambda$ is an eigenvalue, so is $-\lambda$. The
 generic case therefore has the roots of the secular equation for ${\cal M}$ appearing as
 quartets $\alpha,-\alpha,\alpha^{*},-\alpha^{*}$ or as pairs of real eigenvalues $r,-r$.
  To summarize, we may write (the dimension $N$ is 12$V$ for 4D QCD, 2$V$ for
  2D QED, where $V$ is the lattice volume = number of lattice sites)
\begin{equation}
\label{eq:sec}
 {\rm det}(\lambda - {\cal M})= \lambda^N +c_2\lambda^{N-2}+c_4\lambda^{N-4}+...c_N
\end{equation}
 where the $c_i$ are all real and given in terms of ${\cal M}$ by standard trace formulas
\begin{eqnarray}
\label{eq:seccoeff}
  c_2 &=& -\frac{1}{2}{\rm Tr}({\cal M}^2) \\
  c_4 &=& -\frac{1}{4}({\rm Tr}({\cal M}^4)-\frac{1}{2}{\rm Tr}({\cal M}^2)^2) \\
  c_6 &=& -\frac{1}{6}({\rm Tr}({\cal M}^6)+\frac{1}{8}{\rm Tr}({\cal M}^2)^3-\frac{3}{4}{\rm Tr}({\cal M}^2)
{\rm Tr}({\cal M}^4)),\;\;\;{\rm etc}
\end{eqnarray}
 Henceforth we suppose that $r$=1.  The factors $1\pm \gamma_{\mu}$ appearing
 in ${\cal M}$ are then projection operators, and in particular $(1+\gamma_{\mu})(1-\gamma_{\mu})$
 vanishes. Consequently, all terms in ${\rm Tr}({\cal M}^2)$, which necessarily correspond to
 products ${\cal M}_{ij}{\cal M}_{ji}$ where $i,j$ involve nearest neighbour lattice sites, automatically
 vanish. As a result
\begin{equation}
\label{eq:c2zero}
  c_2 = {\rm Tr}({\cal M}^2) =0
\end{equation}
 and the sum of squares of the complex eigenvalues of the Wilson-Dirac matrix must
 vanish in an arbitrary gauge configuration! This sum rule is very useful in checking
 the accuracy of the (sometimes  unstable) numerical routines we have used to
 effect the spectral resolution of ${\cal M}$.  The calculation of $c_4$ is equally
 straightforward. Again, each contributing term in ${\rm Tr}({\cal M}^4)={\cal M}_{ij}{\cal M}_{jk}{\cal M}_{kl}{\cal M}_{li}$
 corrresponds to a  sequence of 4 links forming a closed path and with (for $r$=1)
 backtracks forbidden. The resulting color trace simply produces the plaquette trace
 for an elementary square on the lattice, provided that the linear dimension of the
 lattice (in all directions) is greater than 4, so that Polyakov lines stretching across the
 whole lattice are excluded. In  D=2,4 dimensions, one finds
\begin{eqnarray}
  {\rm Tr}({\cal M}^4) &=& -D\sum_{P}{\rm ReTr}(U_P) \\
  c_4 &=& \frac{D}{4}\sum_{P}{\rm ReTr}(U_P) 
\end{eqnarray}
 which gives the sum rule that the fourth power of eigenvalues of ${\cal M}$ must sum to 
 the gauge action. Similarly, $c_6, c_8,..$ involve sums of non-backtracking 
 gauge-invariant loops
 (and eventually, Polyakov lines) of length 6,8, etc.  Note the symmetry property
 under simultaneous inversion of all links
 (which will be used below in discussing the analytic structure of quenched amplitudes) 
 $c_n(\{U_l\})=c_n(\{U_{l}^{-1}\})$.

    Similar calculations show that the quantity ${\rm Tr}({\cal M}{\cal M}^{\dagger})$ is independent 
 of the gauge field configuration. Namely
\begin{eqnarray}
\label{eq:2Drad}
  {\rm Tr}({\cal M}{\cal M}^{\dagger})&=&4V\;\;\;({\rm 2D\;\;QED})  \\
\label{eq:4Drad}
  {\rm Tr}({\cal M}{\cal M}^{\dagger})&=&16N_{c}V\;\;\;({\rm 4D\;\;QCD})
\end{eqnarray}
This implies, for example, that the spectrum of ${\cal M}$ is rigorously contained within a circle of radius 
 $2L$ for 2 dimensional lattice QED on a $L$x$L$ lattice. A stronger bound follows
 from the observation that (for $r$=1)  ${\rm Tr}{\cal M}^{n}<C\cdot 3^{n}$ for large $n$ as the 
 number of closed non-backtracking paths of length $n$ is clearly $<3^{n}$ in 2
 dimensions and the Dirac traces involve products of projection operators which do
 not grow with $n$. This implies $|\lambda_{a}|<3$ for eigenvalues of ${\cal M}$. 
   In fact with typical gauge
 field configurations,  eigenvalues are not found with magnitude  larger than 2. 
 A typical spectrum for quenched  2D QED (12x12 lattice at $\beta$=4) is shown in Fig(1). The left, right and central branches of
 the Wilson-Dirac spectrum are clearly visible, as well as a pair of  exactly real eigenvalues
 for each branch. 
 One of these real modes has drifted quite a substantial distance into the interior of the
 spectral oval, away from the critical regime. 
\begin{figure}
\psfig{figure=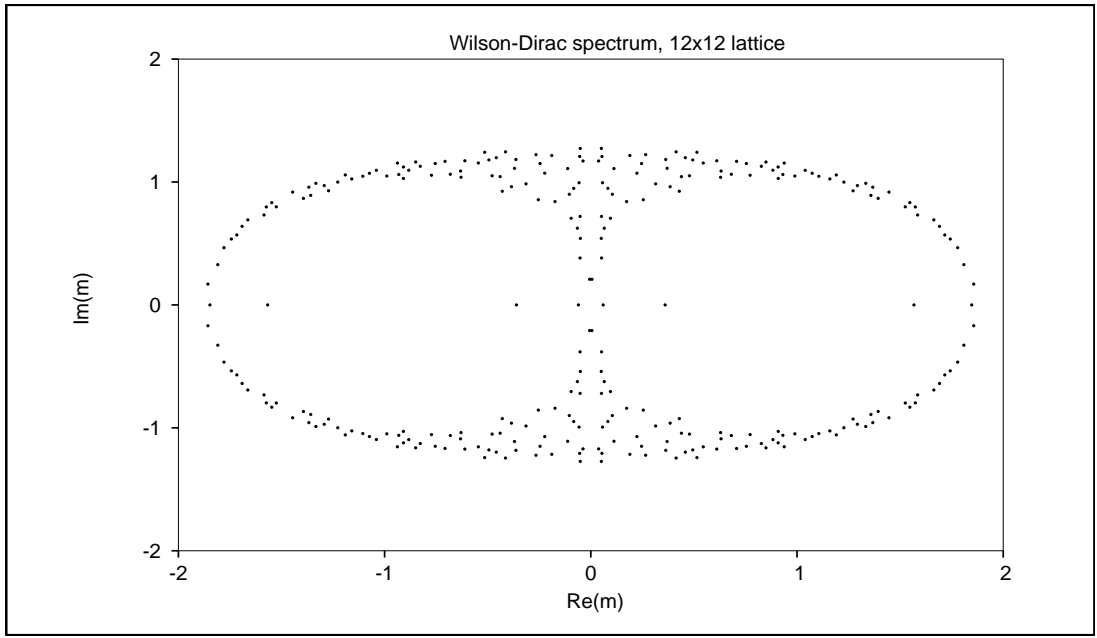,
width=0.95\hsize}
\caption{A typical Wilson-Dirac spectrum in QED2}
\label{fig:spectrum}
\end{figure}

   The discrete symmetries of the Wilson matrix mean that the spectrum is considerably more 
 constrained than would be the case for a generic complex operator. In particular the
 appearance of  exactly real eigenvalues as in Fig(1) is {\em not} a miracle. As a simple toy example,
 consider $N$=4, and assume that the secular equation is constrained, as for our 
 Wilson matrix, to have only even terms with real coefficients. For a general value
 of the $r$ parameter (not equal to unity), we would have $c_2\neq 0$. The resulting  4 roots
  are all real  provided only that  $c_2 <0$, $c_2^2 >c_2^2-4c_4>0$.
  The reality of such roots is clearly robust with respect to small variations of $c_2,c_4$.
 (For $r$=1,  similar arguments show the robustness of the reality of roots of
 the cubic secular equation for $N$=6). 
  Explicit calculations (described below) of the full spectrum in 2D lattice QED on a variety
  of lattices  confirm that  when real eigenvalues  of ${\cal M}$ appear for some set of link angles,   
   they remain real in a
  finite neighborhood of this point in the configuration space of gauge link variables. The 
 appearance of exactly real  modes may seem unsurprising in view of the {\em necessity} 
 of such modes in a topologically nontrivial background in the continuum, but 
 such  modes do not appear in the Kogut-Susskind (KS) formulation of staggered fermions on the
 lattice \cite{KS}, where the fermion matrix (essentially $D$ of (5)) is exactly skew-hermitian, so that
 the spectrum is confined to the imaginary axis, and an exactly zero eigenvalue is  
 nongeneric. Instead, the latticized KS matrix displays pure imaginary
 eigenvalues {\em close}  to, but
 not exactly, zero
 in a smooth background lattice field (e.g. after cooling)  of nonzero winding number.

\section{Mass Singularity Structure of the Quenched Functional Integral}

    The appearance of poles in the physical region for  bare quark masses is a symptom
  of  a serious disease in the standard quenched functional integral.  Strictly speaking,
   this integral - even on a finite lattice,  with compact integrations over gauge link
   variables-  {\em does not exist}  for values of the bare quark mass $m$ in the physical
   range. The quenched functional integral can in principle 
 be defined by analytic continuation from
   a pole-free region (see below)  in the complex plane  of $m$,
 but as we shall soon show  the analytic structure of quenched amplitudes
   becomes extremely complicated in the region of physical interest. In any event  analytic continuation
  is out of the question for  a numerically computed amplitude. We are forced to face the
  unpleasant fact that the quenched functional integral with Wilson fermions, 
 as defined in a manner amenable to Monte Carlo sampling, runs directly
   over poles of the integrand.  Staying at large quark masses would 
 avoid this problem, which only appears, as we shall see, at kappa values
 $\kappa>\kappa_0$, where $\kappa_0$ is the critical kappa value for the
 free theory (=0.25 in 2 dimensions, 0.125 in 4 dimensions). On the other 
 hand, Monte Carlo simulation of the quenched integral 
 at smaller, physically interesting bare quark mass  will  necessarily give
  ill-defined results, even when very large ensembles are used.

   In a full dynamical simulation of hadron correlators in lattice QCD, one computes
 the functional integral over both fermionic and gauge degrees of freedom of a
 product of an even number of quark fields
\begin{eqnarray}
  <\psi_{i_1}\psi_{i_2}..\psi_{i_n}\bar{\psi_{j_1}}\bar{\psi_{j_2}}..\bar{\psi_{j_n}}>
 &=&\frac{1}{Z_0(m)}\int d\psi_{i}d\bar{\psi_{j}}dU_{l}\psi_{i_1}\psi_{i_2}..\psi_{i_n}\bar{\psi_{j_1}}\bar{\psi_{j_2}}..\bar{\psi_{j_n}}   \nonumber \\  
  &&e^{-S_g(U_l)-\bar{\psi_{i}}(m+{\cal M}_{ij})\psi_{j}}  \\
\label{eq:functint}
 &\equiv&\frac{N(m)}{Z_0(m)}
\end{eqnarray}
  Such $2n$-point functions are necessarily meromorphic functions of the bare quark mass
 provided the determinant arising from the Grassmann integration is properly included.
 Indeed,  by the usual rules of Grassmann integration, on a finite 
 lattice both the numerator $N(m)$ and the
 vacuum amplitude $Z_0(m)$ in  (13) are polynomials in the quark mass.
 In the quenched approximation, on the other hand, the absence of the determinantal factor
 results in a much more complicated analytic structure in the complex $m$-plane, as
 poles in quark propagators $D_{ij}\equiv ({\cal M}+m)^{-1}_{ij}$ remain uncancelled by the
 corresponding zeroes in ${\rm det}({\cal M}+m)$. The position of these poles in the 
 complex $m$-plane  is of course
 dependent on the gauge field and consequently the location of singularities in $m$
 of  $N(m)$, where the gauge field integration has been performed, depends on 
 whether the gauge field integrals are prevented from deformation by either a pinch
 or an end point singularity.  (Of course, in the quenched approximation, 
 the normalizing denominator $Z_0(m)$
  only involves the pure gauge action and is consequently 
  independent of $m$, so the entire singularity structure derives from the numerator 
 in (13).)  The generic structure of a quenched contribution to $N(m)$ is thus
\begin{equation}
\label{eq:singint}
  \int \Pi_{l}dU_l  \frac{1}{(m^N +c_2(U_l)m^{N-2}+c_4(U_l)m^{N-4}+..c_N(U_l))^q}\cdot{\rm (analytic\;\;
 in\;\; m, U_{l})}
\end{equation} 
where $q$ is the number of quark propagators appearing in the correlator. Here we are
 examining the analytic structure in the complex plane of a single quark mass- the issue
 of several flavors, and the effect of hairpin graphs in isosinglet sectors in reducing the
 level of singularity in (\ref{eq:singint}) will be discussed below. As the coefficients
 $c_m(U_l)$ are all gauge-invariant, a maximal tree of links may be assumed fixed to
 unity and the integrals in  (\ref{eq:singint}) only include physical degrees of freedom.

   For the rest of this section
 we shall consider abelian gauge theory on a  2 dimensional $L$x$L$
 lattice, although much of
  the discussion applies just as well to 4D QCD.  The gauge integrations are then over link
  angles $0\leq\theta_l\leq2\pi, 1\leq p$ where $p$ is the number of physical (non-fixed)
  links.  From the bound on the magnitude of eigenvalues of ${\cal M}$ implied by
 (\ref{eq:2Drad}), it is clear that  $N(m)$ is analytic for $|m|>2L$, and that singularities
 are confined to a compact central region around the origin of the $m$ plane with poles (if
 any) and paired branch points arranged symmetrically around the origin and real
 axis .  Alternatively, defining $z_l\equiv e^{i\theta_l}$, the link integrations can be thought of
  as over a $p$-dimensional torus $T$, with singularities in $m$ arising whenever two or more
  root surfaces $m^N+c_2(z_l)m^{N-2}+...c_N(z_l)=0$ pinch  $T$ from opposite sides. That
 branch point singularities can easily arise in this fashion is readily seen from a simple
  example (involving only a single physical link) where the integral over the link variable 
 $z\equiv e^{i\theta}$ is written as a complex integral over the unit circle
\begin{equation}
\label{eq:toyint}
   f(m)\equiv\oint\frac{dz}{2\pi i z}\frac{1}{m^4+\alpha(z+1/z)m^2+1}=\frac{1}{\sqrt{(1+m^4)^2-4
\alpha^2m^4}}
\end{equation}
Here $\alpha$ is a real constant. The branch point singularities located at the roots of
 $(1+m^4)^2=4\alpha^2m^4$ arise from a pinch of the unit circle integration contour by
 the two roots in $z$ of the integrand in (\ref{eq:toyint}). 
    More generally, the roots of ${\rm det}(m-{\cal M}(z_{l}))$ define singularity surfaces of the integral
\begin{eqnarray}
\label{eq:qnchint}
  f(m)&=&\oint \frac{1}{{\rm det}(m-{\cal M}(z_l))}\prod_{l}{\frac{dz_l}{2\pi iz_l}} \\
         &=&\oint\frac{1}{m^N+c_2(z_l)m^{N-2}+....}\prod_{l}{\frac{dz_l}{2\pi iz_l}}
\end{eqnarray}
 This integral is well-defined for $|m|>2L$, and defines a real-analytic function of
 $m$ there, but will encounter difficulties as $m$ is brought into the central region
 where roots of ${\rm det}(m-{\cal M}(z_{l}))$ can occur. In particular, branch points
of $f(m)$  in the complex plane of the mass variable $m$ will occur at  points $m_0$
 whenever \cite{ELOP} the hypertorus of integration in (\ref{eq:qnchint}) is pinched from
 opposite sides by two branches of the singularity surface $S_{1}(z_l,m_0)$ and
$S_{2}(z_l,m_0)$ with
\begin{equation}
\label{eq:pnchcond}
   \alpha_1\frac{\partial S_1}{\partial z_l}+\alpha_2\frac{\partial S_2}{\partial z_l}=0
\end{equation}
 for fixed $\alpha_1,\alpha_2$ and all links $z_l$. For example, setting all $z_l=1$
 (the ordered free configuration), there is (for periodic boundary conditions) always
 a pair of real roots at  $m_0=\pm 2$.  If $m_0$ is moved infinitesimally away from these
 points (say to $m_0 = 2\pm \epsilon$) roots of ${\rm det}(m-{\cal M}(z_l))$ appear with
 $|z_{l}| <1$ and $|z_{l}| >1$,  where the configuration $\{z_{l}^{-1}\}$ is necessarily a root
 if $\{z_{l}\}$ is (as $c_n(z_l)=c_n(z_{l}^{-1})$).  Thus the singularity surfaces are naturally
 paired by taking $S_{2}(z_{l},m)\equiv S_{1}(z_{l}^{-1},m)$, in which case (\ref{eq:pnchcond})
 is automatically satisfied with $\alpha_1=\alpha_2$. The surfaces come together and pinch
 the  integration hypertorus at a single point provided $z_{l}=z_{l}^{-1}$ for all links, 
 i.e. if $z_{l}=\pm 1$. We conclude that branch points of $f(m)$ are present at all
 complex values of $m$ which solve ${\rm det}(m-{\cal M}(z_{l}))=0$ for any configuration where
 the link variables are either 1 or -1. In particular, the eigenvalues of ${\cal M}$ for the ordered 
 configuration are all branch points of the quenched amplitudes. In addition, there are
 many other branch points corresponding to configurations where the link variables are
 allowed to take the values $\pm 1$. A glance at Fig(3), which shows branch points arising from
 the ordered configuration as well as from a disordered configuration where the link
 variables were set to $\pm 1$ randomly (on a 6x6 lattice), 
indicates just how complicated the analytic
 structure of a  lattice quenched amplitude can be. In particular, branch cuts extend along
 the entire real axis from $m=-2$ to $m=2$, exactly in the region where one attempts to 
 approach continuum behavior. 

\begin{figure}
\psfig{figure=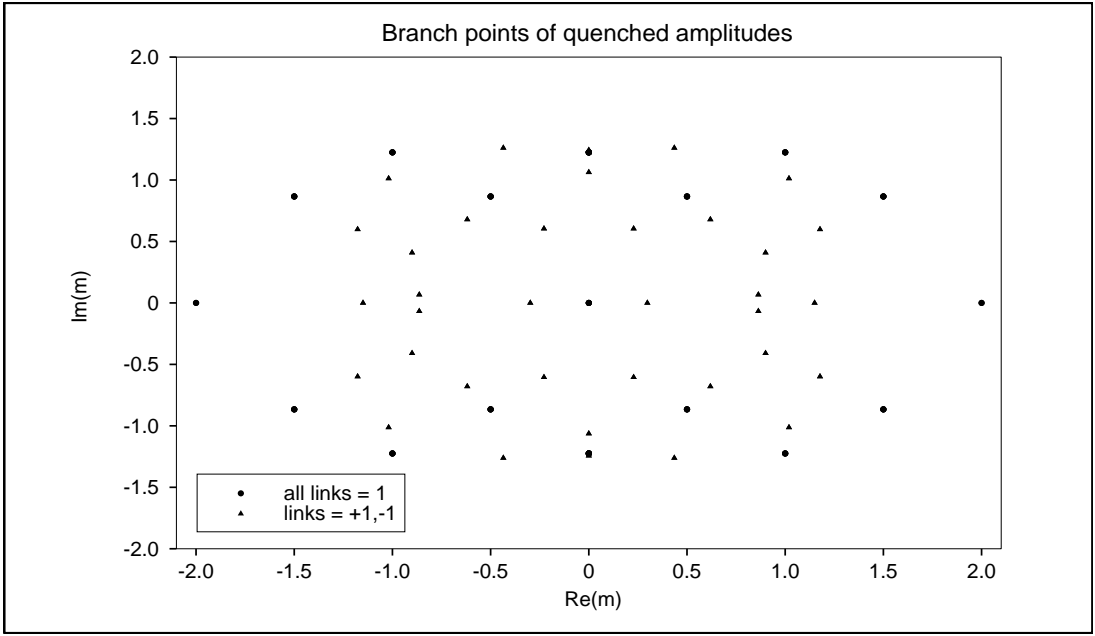,
width=0.95\hsize}
\caption{Location of some of the branch points for L=6}
\label{fig:branch_cuts}
\end{figure}

\newpage
\section{Zero modes, topological charge, and cooling}

  As discussed above in Section 2, there is a clear connection in the continuum theory
 between the topological charge of the background gauge configuration and the zero
 mode structure of the Dirac operator. On the lattice, this connection is necessarily 
 less precise, as discretization effects  smear out both the shape of the Dirac spectrum
 and alter the value of the topological charge associated with a given lattice gauge
 field. As discussed by Smit and Vink, two possible definitions of topological charge
 for 2D QED on the lattice are
\begin{equation}
\label{eq:q1def}
  Q_{1} \equiv\frac{1}{2\pi}\sum_{P}\sin{(\theta_{P})}
\end{equation}
 where $\theta_{P}$ is the plaquette angle for plaquette $P$. This definition
 of course gives in general a noninteger value for the topological charge. 
 Alternatively, one may define
\begin{equation}
 \label{eq:q2def}
 Q_{2}\equiv \frac{1}{2\pi}\sum_{P}\theta_{P},\;\;\;-\pi<\theta_{P}\leq \pi
\end{equation}
 which gives an integer value for the topological charge, agreeing with
 (\ref{eq:q1def}) for smooth fields in the continuum limit.  For a typical
 gauge configuration at weaker values of the coupling the
 number of exact zero modes found for the lattice Wilson-Dirac operator
 agrees with $Q_{2}$ (in each critical branch), with $Q_{2}$
 essentially the closest integer to $Q_{1}$, while at lower $\beta$ one
 finds increasingly frequent
  discrepancies in these three measures of topological charge. 

   A convenient way of understanding the role of discretization on a given 
 finite lattice involves  cooling a Monte Carlo generated configuration to gradually
 remove high-momentum components from the gauge configuration. Such cooling
 tends to stabilize any nontrivial topology while bringing the zero mode and
 topological properties of the system closer to continuum-like behavior.  A typical
 result of a series of cooling sweeps (in which the action is adiabatically lowered,
 effectively by running the simulation at a high fixed $\beta$ value) on the near-real-axis
 spectrum of  ${\cal M}$ (left branch) is shown in Table(1). As the initial configuration,
 with topological charge $Q_1$=0.85 is cooled, the action decreases steadily, the topological
 charge $Q_1$ increases towards unity, and the norm fraction of the positive chirality 
 part of the corresponding real eigenmode increases rapidly towards 1.  There is
 a single exactly real eigenmode (in each branch) of the Wilson-Dirac matrix throughout. 
 Restricting our attention to the left critical branch only, cooling
 moves the real mode gradually further to the left (as we would expect by forcing the system to
 a larger effective $\beta$), but leaves the number of real modes unchanged. Note that the 
 discrete definition $Q_2$ gives exactly unity for the topological charge throughout.

\begin{table}
\renewcommand{\baselinestretch}{1.0}
\centering
\caption{Cooling a configuration with topological charge = 1} 
\vspace{.1in}
\label{tab:cool}
\begin{tabular}{|c|c|c|c|c|c|}
\hline
\multicolumn{1}{|c|}{Cooling sweeps}
&\multicolumn{1}{c|}{ $Q_{1}$}
&\multicolumn{1}{c|}{ $Q_{2}$ }
&\multicolumn{1}{c|}{ Action}
&\multicolumn{1}{c|}{ Real modes (left branch)}
&\multicolumn{1}{c|}{Chirality Fraction} \\[2pt]
\hline
0  & 0.850  & 1 & 20.8 & -1.892 & 0.930  \\
1  & 0.913  & 1 & 14.2 & -1.901 & 0.999\\
2  & 0.940 &  1 & 10.1 & -1.911 & 0.999 \\
3  & 0.950 & 1   & 6.9 & -1.919  & 0.999\\
4  &  0.959 & 1 & 4.7 & -1.926 & 0.999\\
\hline
\end{tabular}
\end{table}

  In the course of a quenched Monte Carlo simulation topological charge fluctuations
 arise resulting in the appearance of new exactly real eigenmodes of the Dirac
 operator. An interesting example of this phenomenon is illustrated in 
 Fig.3.   Initially (circular points)
 there is one exactly real eigenvalue per branch, and a pair of slightly off-real-axis
 complex modes considerably to the right of the left-branch critical line, which for the lattice in
 question (12x12 at $\beta$=5) is at  Re(m)$\approx -1.91$.  Both $Q_{1}$ and $Q_{2}$
 give values for this initial Monte Carlo configuration close to  +1 ($Q_{2}$ is of course
 exactly 1), as shown in Table(2).  The  corresponding  eigenvector would have 
 exactly positive chirality in the continuum- instead the norm fraction of the positive
 chirality piece is initially about 98.8\% (Table(2), column 6).  As the cooling proceeds,
 the complex  pair  in Fig(3) move onto the real axis (diamond points in Fig(3)) and then
 split laterally, with one member moving out towards the critical line.  
 The topological charge $Q_{1}$ moves smoothly up
 towards 2 as this happens, with the discrete version $Q_{2}$ switching from 1 to 2 
 at roughly the point when the second exactly real mode appears (we take snapshots
 of the entire spectrum after each cooling sweep through the whole lattice, so it is
 difficult to be more precise as to the exact point at which the new real modes appear).
 Note that the ``right chirality" fraction of the new zero mode is initially rather low (less
 than one-half), but increases steadily as this mode becomes stabilized and moves
 leftward into the critical region.

\begin{table}
\renewcommand{\baselinestretch}{1.0}
\centering
\caption{Effect of cooling on a topological charge fluctuation } 
\vspace{.1in}
\label{tab:cool}
\begin{tabular}{|c|c|c|c|c|c|}
\hline
\multicolumn{1}{|c|}{Cooling sweeps}
&\multicolumn{1}{c|}{ $Q_{1}$}
&\multicolumn{1}{c|}{ $Q_{2}$ }
&\multicolumn{1}{c|}{ Action}
&\multicolumn{1}{c|}{ Real modes (left branch)}
&\multicolumn{1}{c|}{Chirality Fraction} \\[2pt]
\hline
0  & 1.167  & 1 & 42.4 & -1.90 & 0.988  \\
2  & 1.266  & 1 & 25.9 & -1.92 & 0.994\\
4  & 1.486 &  2 & 15.7 & -1.94,-1.71 & 0.995,0.462 \\
6  & 1.636 & 2   & 10.3 & -1.95,-1.82  & 0.995,0.673\\
8  &  1.693 & 2 & 7.45 & -1.96,-1.86 & 0.995,0.771\\
10 &  1.817 & 2 & 5.33 & -1.96,-1.90 & 0.997,0.897\\
\hline
\end{tabular}
\end{table}

\begin{figure}[htb]
\psfig{figure=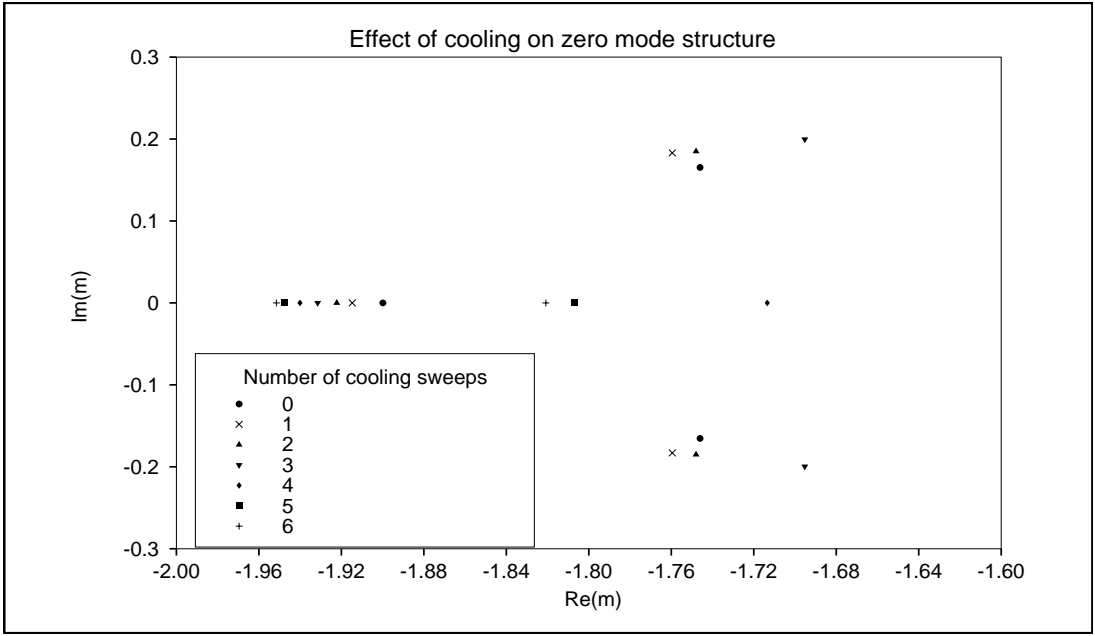,
width=0.95\hsize}
\caption{Effect of cooling on near-zeromodes ($\beta$=5,12x12 lattice)}
\end{figure}
 Evidently, in this case the initial configuration, with $Q_{1}, Q_{2}$ of order unity, actually contains
 a ``nascent" topological charge 2 structure, with large high momentum fluctuations
 hiding the larger value of topological charge. The presence of this hidden topological
 charge is nevertheless signalled by the presence of  complex pairs in the interior of
 the spectral oval, close to the critical region. In the case illustrated here, the effect of
 cooling mimics the further evolution under Monte Carlo updating, which also leads
 to $Q_{1}\approx 2$ after a few more sweeps, with a spectrum containing two exactly
 real modes per branch.  In other cases, we have observed a reduction of topological
 charge under cooling: a pair of real eigenvalues in the left and central branches 
 move out into the interior of the spectral oval, colliding there, then moving
 vertically away from the real axis, eventually to join the noncritical modes far from the
 real axis.

\newpage

\section{Lattice Volume and Lattice Spacing Dependence of Real Spectrum}
    It has become part of the lore of quenched lattice QCD simulations that the
 exceptional configuration problem appears to be ameliorated at fixed $\beta$
 (i.e. lattice spacing)  when one goes to larger physical volumes. It is therefore
 of some interest to see whether this feature can be understood in terms of the
 volume dependence of the real part of the Wilson-Dirac spectrum. As one increases
 the physical volume, one naturally expects more frequent occurrence of configurations
 with nonvanishing topological charge, so it is clear that the overall number of  exactly
 real zero modes, which as we have seen above are clearly correlated with
 nontrivial topological structures, should increase with volume.  The appearance
 of exceptional configurations requires however that these modes appear in the
 quark mass regime being studied, just to the left of the critical line corresponding to 
 zero renormalized mass.  

    The simulations described in the preceding section show that the number of exact
 zero modes is roughly  correlated with the absolute magnitude of the topological charge
 $Q_{1}\equiv\frac{1}{2\pi}\sum_{P}\sin{(\theta_{P})}$.  Quenched compact 2 dimensional QED
 is an exactly solvable model in the pure gauge sector, so it is a straightforward exercise
 to evaluate the distribution of this quantity over quenched configurations generated on
 a LxL lattice (periodic boundary conditions) at any given $\beta$. If the lattice volume
 is sufficiently large that $(I_{1}(\beta)/I_{0}(\beta))^{L^{2}}$ (which decreases
 exponentially fast with lattice volume)  is negligible, where $I_{0},
 I_{1}$ are the usual modified Bessel functions, then periodic and free boundary 
 conditions become equivalent  and the distribution of $Q_{1}$ becomes exactly
 Gaussian.  (For the smallest lattice used below, L=6 at $\beta$=4, this approximation incurs
 an error of  order 10$^{-3}$). One finds
\begin{equation}
\label{eq:topdist}
  \rho(Q_{1})=Ce^{-2\pi^{2}\frac{\beta I_{0}(\beta)}{I_{1}(\beta)}\frac{Q_{1}^{2}}{L^{2}}}
\end{equation}
 which implies that the average value $<|Q_{1}|>$ rises linearly with the lattice size, or
 as the square root of the lattice volume. We have evaluated the spectra for 2500 
 configurations of lattices at $\beta=$4 and sizes L=6,8, and 10. The total number of
 exactly real  modes rises in this range of lattice sizes somewhat faster than L though not
 as rapidly as L$^{2}$ - specifically, the average number of exactly zero modes
 per branch was found to be 0.26, 0.47 and 0.63 for the L=6,8, and 10 lattices respectively.
 The spectral histograms for these three cases in the region of the left critical branch 
 is shown in Fig(4).

\begin{figure}[htb]
\psfig{figure=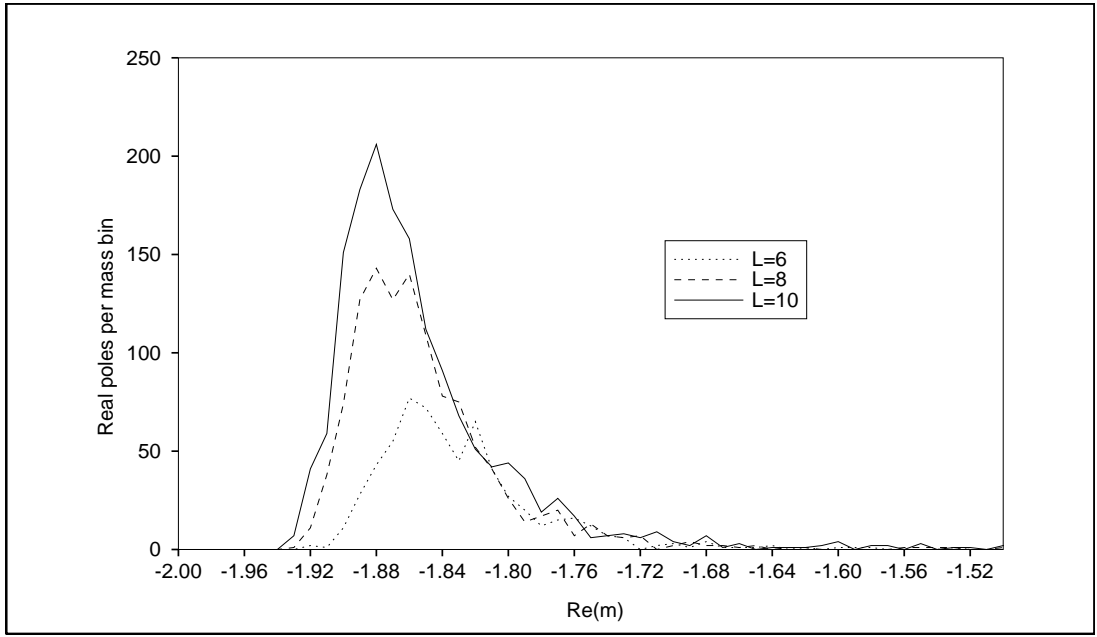,
width=0.95\hsize}
\caption{Volume dependence of spectrum (left branch))}
\end{figure}

  Fig(4) provides a possible clue to the observation that exceptional configurations become 
 rarer at fixed quark mass as the volume increases. We see that although the total number
 of real modes is certainly increasing with lattice volume, the leftmost zero mode shifts
 towards the boundary of the spectral region at $m=-2$ rather slowly (for $L$=6,8, and 10, the
 leftmost modes were  found at  -1.920,-1.923,  and -1.927 respectively), while the 
 peak of the histogram, which we may identify with the critical mass value, shifts leftward
   more rapidly (for $L$=6,8 and 10, the  histograms peak at -1.855, -1.87, -1.88), so that the
  probability of encountering an exceptional configuration as one approaches the 
  critical line from the left decreases with increasing volume if one keeps a fixed offset
 from the critical line to maintain a fixed physical quark mass.  However, these histograms
 also make it clear that the exceptional configurations will necessarily appear at any volume
 once one goes sufficiently close to the critical point.  As we shall now see, the benefits
 to be obtained from increasing lattice volume are far less dramatic than from the reduction
 of lattice spacing (at fixed physical volume). 

\begin{figure}
\psfig{figure=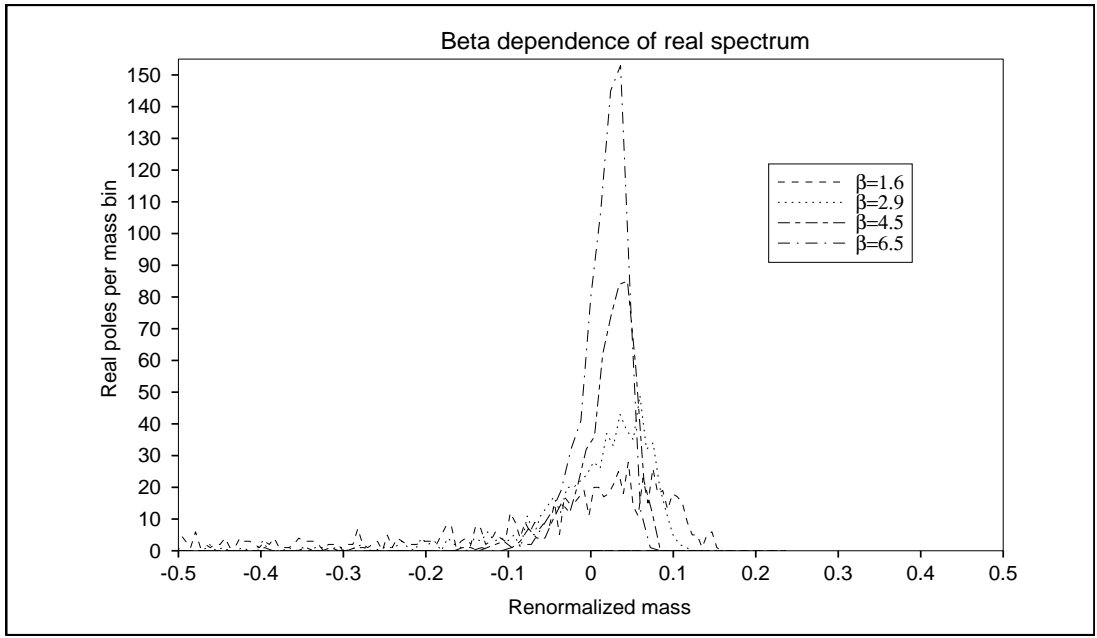,
width=0.95\hsize}
\caption{Beta dependence of spectral histogram}
\label{fig:betahistograms}
\end{figure}

   In Fig(5) we show a histogram of exactly real eigenmodes appearing in quenched
 simulations in QED2 on  $L$x$L$ lattices with $L$=6,8,10 and 12, this time
 with the gauge coupling
 readjusted to keep the physical lattice volume fixed (specifically, we have used 
 $\beta$=1.6, 2.9, 4.5  and 6.5).  The histograms were obtained from 1000 configurations
 separated by 200 quenched Monte Carlo sweeps (except for $L$=12, where measurements were
 separated by 500 sweeps). Note that the direction of the horizontal axis in these plots
 is reversed from the previous spectral plots, as by convention positive quark mass
 corresponds to the region to the {\em left} of the critical line (at roughly 2-0.65/$\beta$).
 The density of real modes near the left critical branch is
 shown, relative to a critical line defined in this case by the peak of the distribution (this
 turns out to be close to but slightly offset from the Smit-Vink critical line value, with the
 offset decreasing with increasing $\beta$), and with
 the scale on the horizontal axis rescaled to keep  physical masses fixed. 
 It is apparent that the spread of dangerous poles into the physical mass region
 becomes acute at strong coupling, and that (at least in 2D QED) the probability of
 encountering exceptional configurations at a fixed renormalized quark mass 
 decreases rapidly as $\beta$ is increased. 

    The manifest asymmetry of the real mode distribution around the critical line accounts
 for the uncontrollably noisy results obtained if one attempts a quenched simulation at lighter
 quark masses approaching from the interior of the spectral oval (i.e. at values of the
 kappa parameter greater than kappa critical). Nevertheless, we shall see below that the 
 modified quenched approximation approach \cite{qzpaper} to resolving the exceptional configurations
 in the $\kappa<\kappa_{c}$ regime is equally successful in removing this noise and
 restoring sensible correlators for $\kappa>\kappa_{c}$.

\vfill\eject

\section{Effects of clover improvement on the Wilson-Dirac Spectrum}

    The dispersion of real axis eigenvalues of the Wilson-Dirac matrix was seen
 above to decrease rapidly with increasing $\beta$, and it is clear that we are dealing with
  a pathology which is directly traceable to the particular discretization of the theory
  implied by an unimproved Wilson action for the quark fields. At first sight this would suggest
  that the problem might be removable (or at least, substantially reduced) by adopting
  a locally improved action, say by adding a clover-leaf term with appropriate coefficient.
  Unfortunately (see also \cite{exceptconf}) this does not seem to be the case. In Fig.6 we
  show the effect of  introducing such a term, with various values of the coefficient, on
  the spectral histogram of real eigenvalues for a 8x8 lattice at $\beta$=4.5 in QED2. 
 Evidently, the clover
 improvement term is simply not effective in restraining the tendency of real modes to
 drift away from the critical region at strong coupling.

\begin{figure}
\psfig{figure=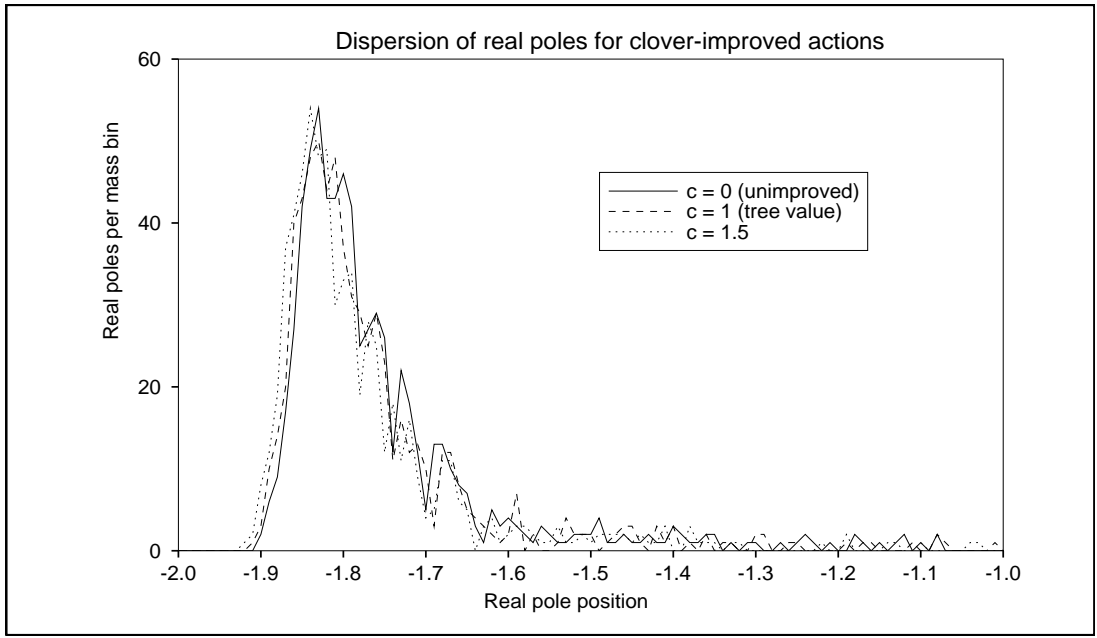,
width=0.95\hsize}
\caption{Dependence of spectral histogram on clover improvement coefficient ($\beta=$4.5)}
\label{fig:clover}
\end{figure}
   
  Of course, the addition of a clover term certainly will shift the location of  real modes
 on each individual configuration. Such a shift can have a dramatic effect on a given
 exceptional configuration, as it may be enough to move the propagator pole far 
 enough from the  chosen quark mass value to remove the large fluctuation in the correlators 
 {\em for that configuration}. But in a statistical sense the problem remains.  Our QED2
 studies have revealed a rather interesting effect  at the individual configuration level:
 it appears that the effect of a clover term is much stronger on the two central branches
 ({\em not} used in typical simulations) of the Wilson-Dirac spectrum.  This effect is shown
 in Fig(7), where the full spectrum on a typical topological charge $Q_{2}=$2 configuration is
 shown for clover coefficient values of  0,1.2, and 2.4 (where the tree value is unity).
 At higher values of the clover coefficient the real modes in the right central branch
 become degenerate and the nonzero modes line up quite accurately with the ones
 on the real axis.  There is also a tendency for the two central branches to separate
 as the clover coefficient is made larger. The effect on the left and right lateral branches 
is less pronounced-
 the two zero modes on the left branch move somewhat apart, while those on the right move 
 together. The whole process
  is reversed left to right if the sign of the clover coefficient is changed.

\begin{figure}[htp]
\vspace{-1.2in}
$$\psfig{figure=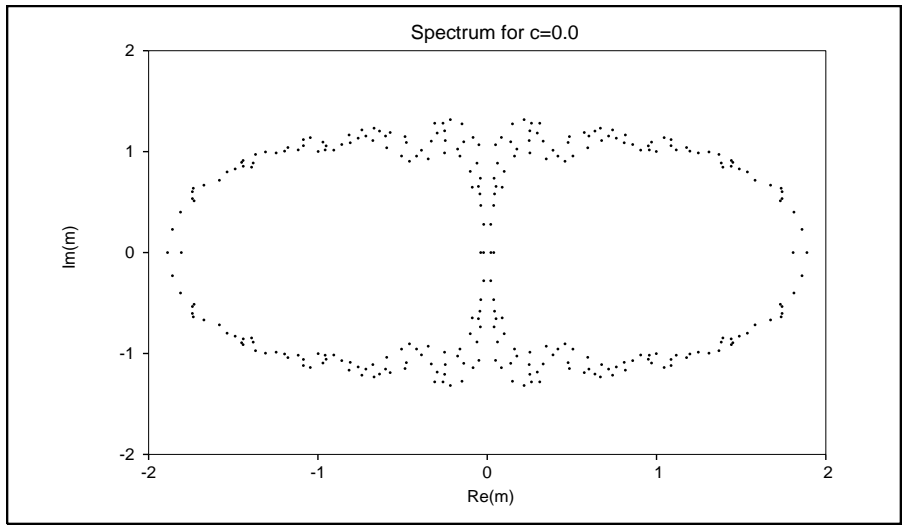,
width=0.78\hsize}$$
$$\psfig{figure=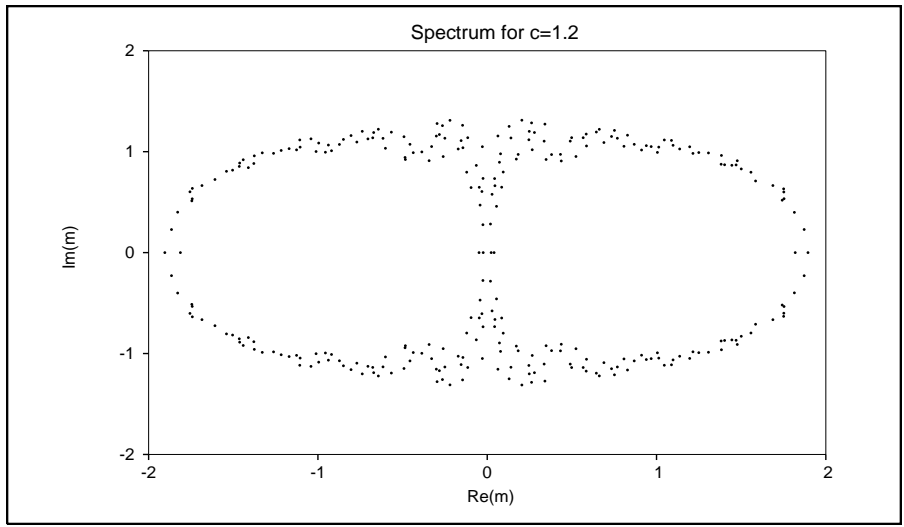,
width=0.78\hsize}$$
$$\psfig{figure=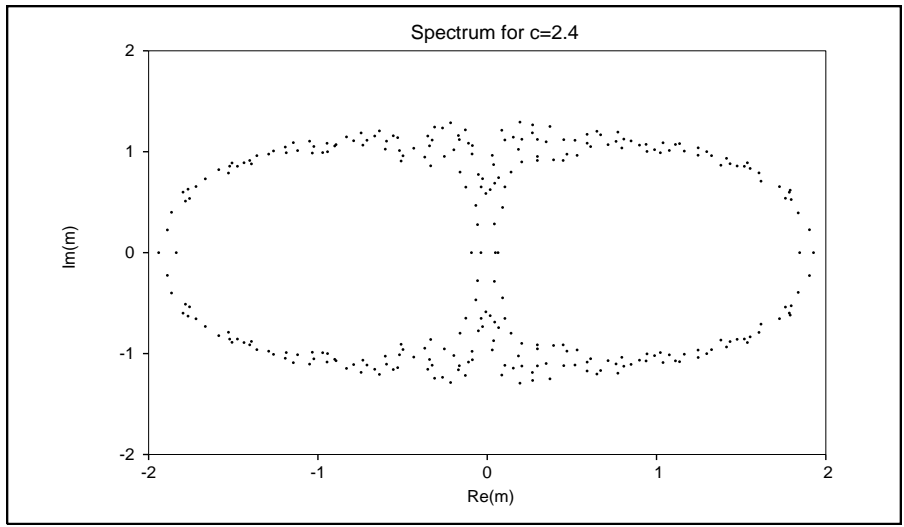,
width=0.78\hsize}$$
\caption{Spectrum of topological charge 2 configuration, clover coefficients =0,1.2,2.4}
\label{fig:cloverc2.4}
\end{figure}

\newpage

\section{Testing the modified quenched approximation}

   Recently, we have proposed  a nonlocal improvement procedure which should become
 arbitrarily precise in the continuum limit, as it involves explicitly shifting the position of
 real poles of the quark propagator by amounts which go to zero in this limit, but in such
 a way as to preserve the location of the critical point to leading order as the lattice
 spacing goes to zero.  A redefinition of the Wilson-Dirac matrix which involves a shift
 in the location of real eigenvalues while leaving the corresponding eigenmodes 
  unchanged  evidently corresponds to a nonlocal addition to the action, as the
  typical eigenmodes for a real eigenvalue will be associated with a topological charge
  fluctuation which extends over several lattice sites  (with power rather than exponential
  falloff away from the ``instanton center"). Of course, we need to be sure that this redefinition
  does not introduce an unwanted sizeable systematic distortion of the quenched averages
  at finite lattice spacing- in particular, does the proposed redefinition,
 which is really a prerequisite to having any precise notion of a quenched 
 lattice amplitude, actually yield results   which are less noisy {\em and}
  closer to the results of a full dynamical simulation on the same lattice?  Direct evidence
  for this will now be presented.

    The modified quenched approximation (MQA) recently proposed by us yields a
 well-defined regularization of the quenched functional integral which is constructed 
 to reduce to the usual version in the continuum limit, in the sense that real axis poles
 of the Wilson-Dirac propagator which have strayed into the physical region for the 
 quark masses are shifted back to the critical line where they would sit in this limit.
 The shift is performed in such a way as to preserve the location of the critical line in
 an average sense (for more details, see \cite{qzpaper}). In the context of QED2, we define the
 MQA modified propagator as follows.  The fermion propagator
 $S\equiv (\frac{1}{2\kappa}+\cal M)^{-1}$ (with the bare quark mass defined as $m_{0}\equiv \frac{1}{2\kappa}
-\frac{1}{2\kappa_{c}}$ in terms of the usual kappa parameters, and  with
 $\frac{1}{2\kappa_{c}}\sim $2.0-0.65/$\beta\equiv\lambda_{c}$ in QED2)
 has the spectral decomposition
\begin{equation}
 S_{ij} =\sum_{a}\frac{\chi_{ai}\tilde{\chi}_{aj}}{m_{0}+\lambda_{c}+\lambda_{a}}
\end{equation}
where $\lambda_{a}$ are the eigenvalues of ${\cal M}$, and the indices
 $i,j$ incorporate spatial and Dirac information. The left-shifted real eigenvalues
 will cause poles at values of $m_{0}$ corresponding to $\kappa<\kappa_{c}$. 
  For QED2 we are able to perform a full spectral resolution for each configuration
. Define  $u \equiv 2m_{0} =\frac{1}{\kappa}-\frac{1}{\kappa_{c}}$
 and replace any pole appearing at a position $u = u_{\rm pole}$ as follows
\begin{equation}
  \frac{1}{u-u_{\rm pole}}\rightarrow \frac{2}{u}-\frac{1}{u+u_{\rm pole}}
\end{equation}
At large mass (or large $u$) the first two terms in the expansion in $1/u$ are identical
 and terms linear in the shifts should average to zero. This procedure preserves the location of $\kappa_{c}$ to first order
 in the dispersion of real eigenmodes around the critical line (which is of course going
 to zero in the continuum limit). 
   The full MQA propagator may be simply computed by adding a term to the naive fermion
 propagator $S_{ij}$ which incorporates the pole shift. Namely, for any configuration where a
  pole(s) appears in the physical region (operationally we have defined this in this paper 
 to be the region between 0 and $2m_{0}$ on the real bare quark mass axis),  we replace
\begin{equation}
 S^{\rm MQA}_{ij}\equiv S_{ij}+A^{\rm pole}(\kappa){\rm Res}^{\rm pole}_{ij}
\end{equation}
with
\begin{eqnarray}
A^{\rm pole}(\kappa)&\equiv&\frac{2}{u}-\frac{1}{u+u_{\rm pole}} -\frac{1}{u-u_{\rm pole}} \\
 {\rm Res}^{\rm pole}_{ij}&\equiv& {\rm lim}_{u\rightarrow u_{\rm pole}}(u-u_{\rm pole})S_{ij}
\end{eqnarray}

   Although the procedure outlined here is certainly valid in the continuum limit, where we
 expect the frequency of exceptional configurations at any fixed physical  quark mass
 to vanish, it is perhaps not apparent that the modification induced by the pole shift
 procedure does not introduce some large and misleading distortion of
 the correlators at stronger coupling.   The level of numerical control possible with
 QED2 allows a direct comparison of the naive quenched, shifted pole, and full dynamical
 results for the interesting correlators of the theory. We have measured the 
 pseudoscalar correlator $<\bar{\psi}(x)\gamma_{5}\psi(x)\bar{\psi}(0)\gamma_{5}\psi(0)>$
 at fairly strong coupling $\beta=$4.5 on a 10x10 lattice for these three situations for
 a variety of  quark masses close to the critical point of the left critical branch.  The
 full dynamical simulations were done using an exact update algorithm \cite{DRV} where the
 determinant and fermion propagator are known exactly at every stage. Typically, 
 measurements were performed after every sweep for 400 sweeps, and the statistical
 error computed by measuring  an autocorrelation time for each correlator  (the
 naive quenched correlators decorrelate very rapidly so autocorrelation times are on
 the order of 1 sweep in the unshifted quenched case, while for the shifted pole and dynamical correlators, the autocorrelation times 
 range from 3-10 sweeps).  QED2 is a superrenormalizable theory with a finite bare
 coupling which fixes the mass scale of the theory, so a direct comparison of quenched
 and dynamical results can be made in this theory without  the need for a compensating
 scale change. As we work on fairly small lattices, it is difficult to find  mass
 plateaus suitable for extracting a meson mass, 
 and we have decided to  compare directly the pseudoscalar correlators, which
 are after all physical quantities with unambiguous values (for given $\beta$
 and lattice size) in the full unquenched theory.

    The comparison of naive quenched, MQA and full dynamical simulations for quark mass
 values 0.08, 0.10 and 0.115 are shown in Fig(8). The quark mass values quoted here
 indicate the offset from a critical value on a 10x10 lattice
 at $\beta$=4.5 defined as the average real part of the left branch real eigenmodes
 with $-2 < {\rm Re}(\lambda)<-$1.5 .  The increase in the naive quenched errors at
 smaller quark mass 
 can be directly traced to the increasing frequency of nearby real poles. In all cases, the
 MQA correlator is less noisy and {\em closer} to the full dynamical result. Statistical 
 errors in the latter are related to autocorrelation problems, not the intrinsic noisiness
 of the data.  Of course, for very small quark masses, we begin to see a deviation
 between MQA and full dynamical results, as the MQA does not contain the suppression of
 nontrivial topologies implicit in the zero mass full dynamical theory. 

\begin{figure}[htp]
\vspace{-0.8in}
$$\psfig{figure=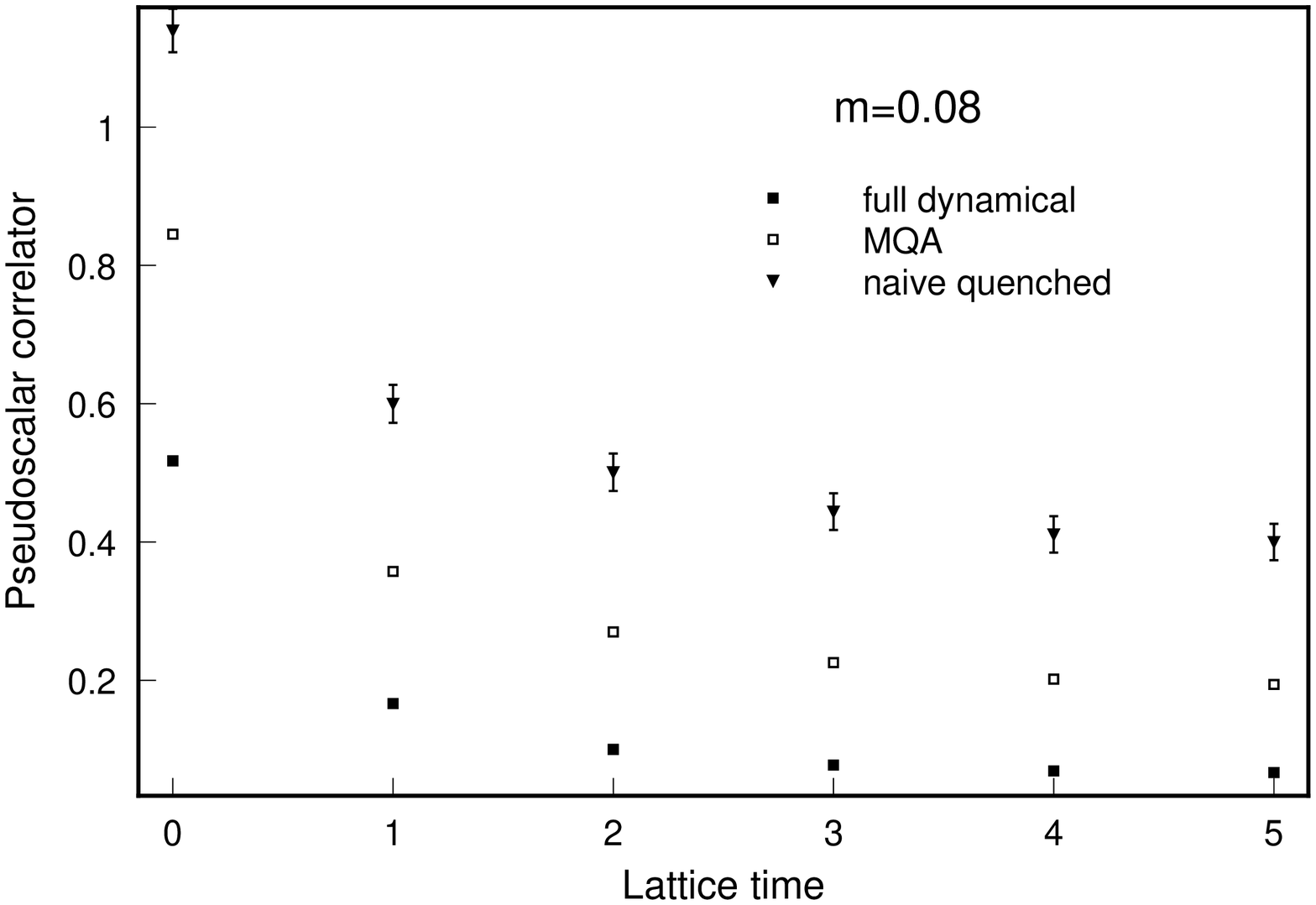,
width=0.74\hsize}$$
$$\psfig{figure=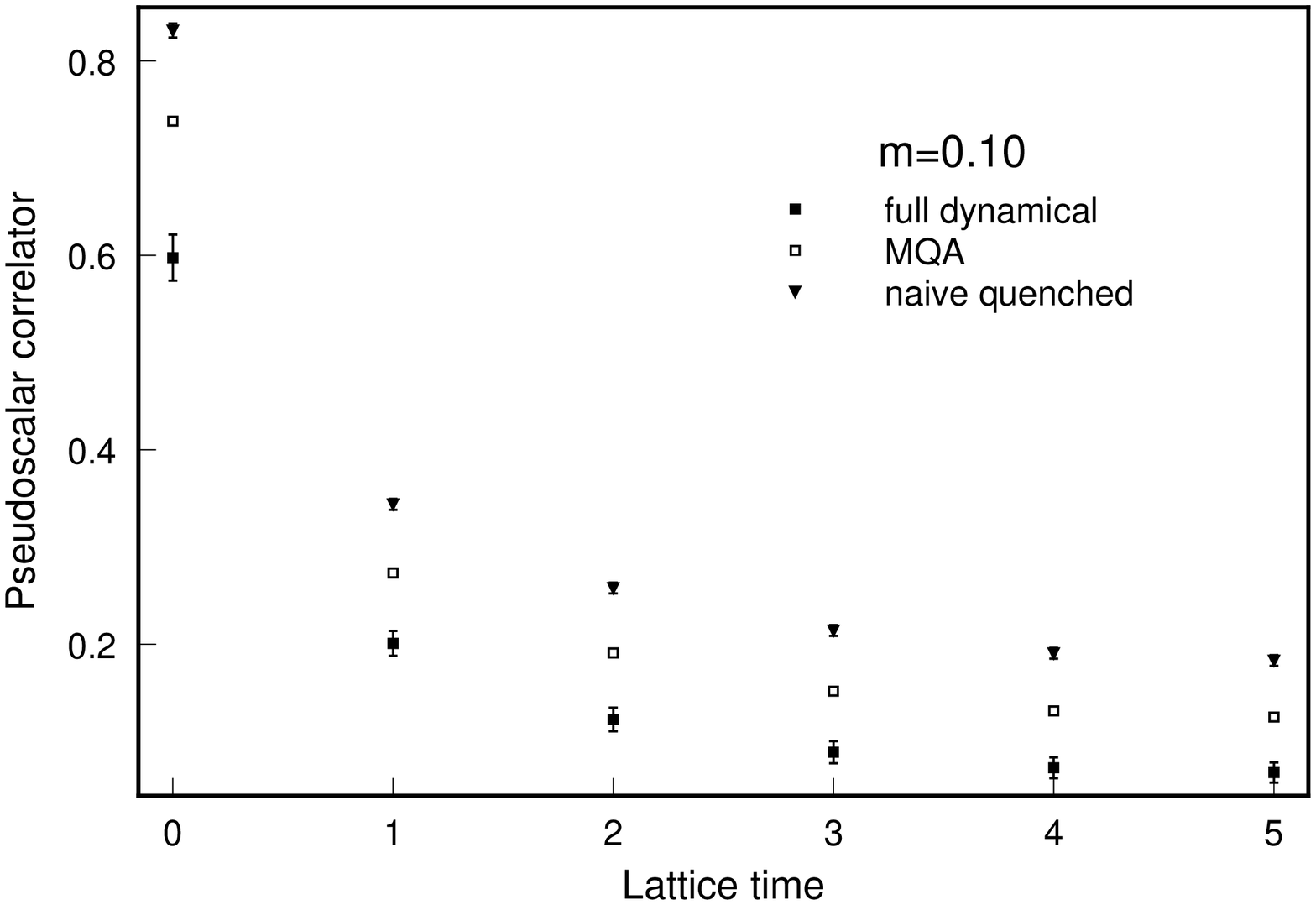,
width=0.74\hsize}$$
$$\psfig{figure=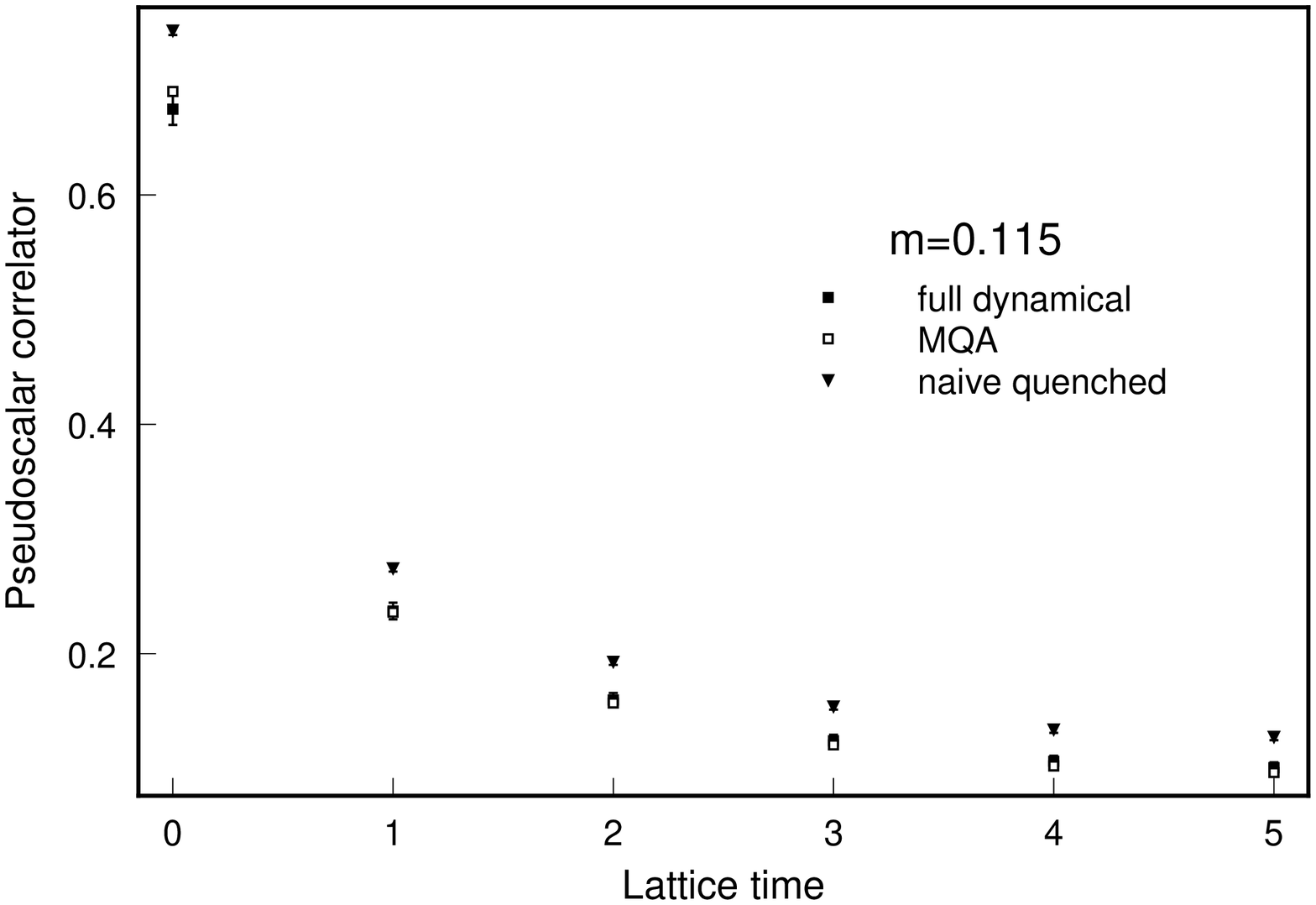,
width=0.74\hsize}$$
\caption{Pseudoscalar correlators in QED2- quenched, MQA and full dynamical}

\end{figure}

Amazingly, the MQA procedure makes it possible (to our knowledge, for the first time) 
 to obtain sensible results by approaching the critical line from the {\rm interior of the spectral
 oval, corresponding to negative quark mass with the conventional  definition (i.e.
 the ``supercritical" regime with $\kappa>\kappa_{c}$). In the
 continuum, a negative quark mass may be removed by a $\gamma_{5}$ redefinition of
 the quark field, but on the lattice this region contains many more real modes extending 
 farther away from the critical line (see Figs. 4,5,6),
 leading  to extremely noisy quenched correlators (although of course the full dynamical
 theory is perfectly well-defined here). The plots shown in Fig(9) for the two point 
function of $\bar{\psi}\gamma_{5}\psi$ show that the MQA
 is effective in this case also, although the increased density of real modes in
 this regime makes the MQA less effective at extremely small quark masses- the
 results shown are for quark masses of -0.15, -0.16  and -0.185 (as before, on a 10x10
 lattice at $\beta$=4.5), with the negative sign
 indicating that we are now working in the interior of the spectral oval . 
 For small quark masses, the results in the naive quenched model are again 
 dominated by noise, while the MQA and full dynamical results are close , with
 the statistical noise under control in the MQA even for quite small {\em negative} quark masses. For larger
 quark masses, where  sensible results can be extracted from an uncorrected quenched
 calculation, the MQA correlators lie between the naive quenched and full dynamical
 results, as was the case previously in the subcritical domain.

\begin{figure}[htp]
\vspace{-0.9in}
$$\psfig{figure=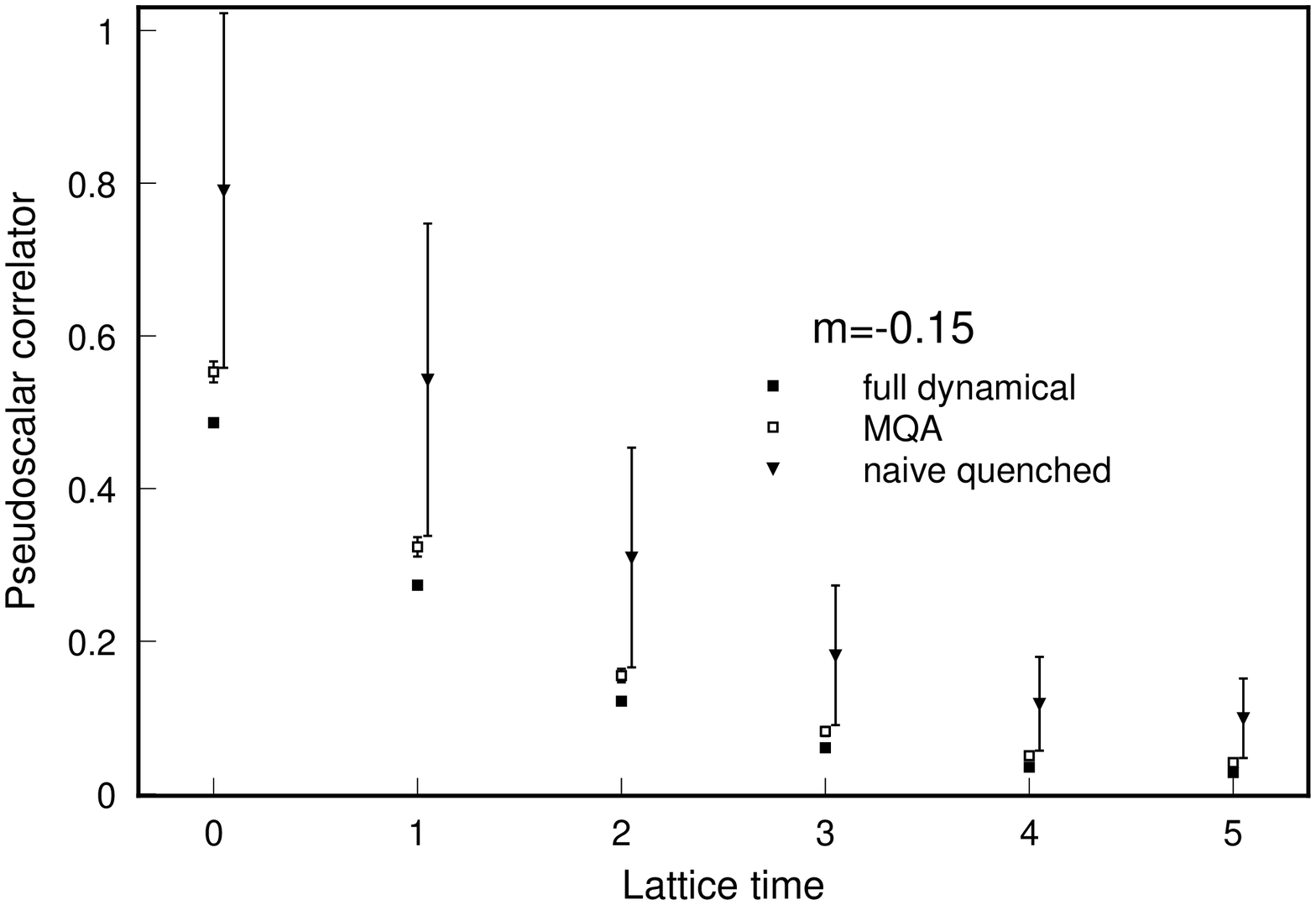,
width=0.75\hsize}$$
$$\psfig{figure=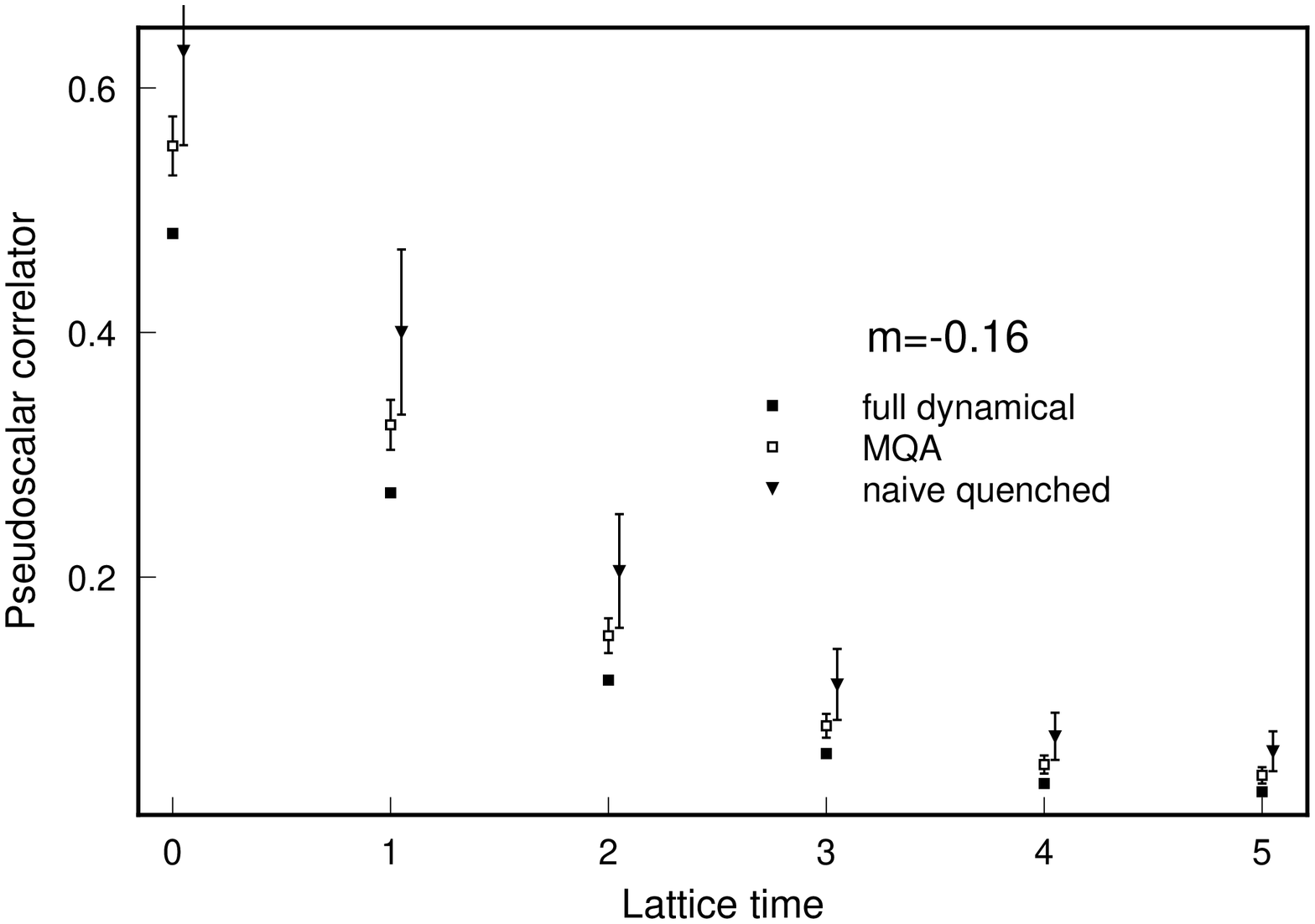,
width=0.75\hsize}$$
$$\psfig{figure=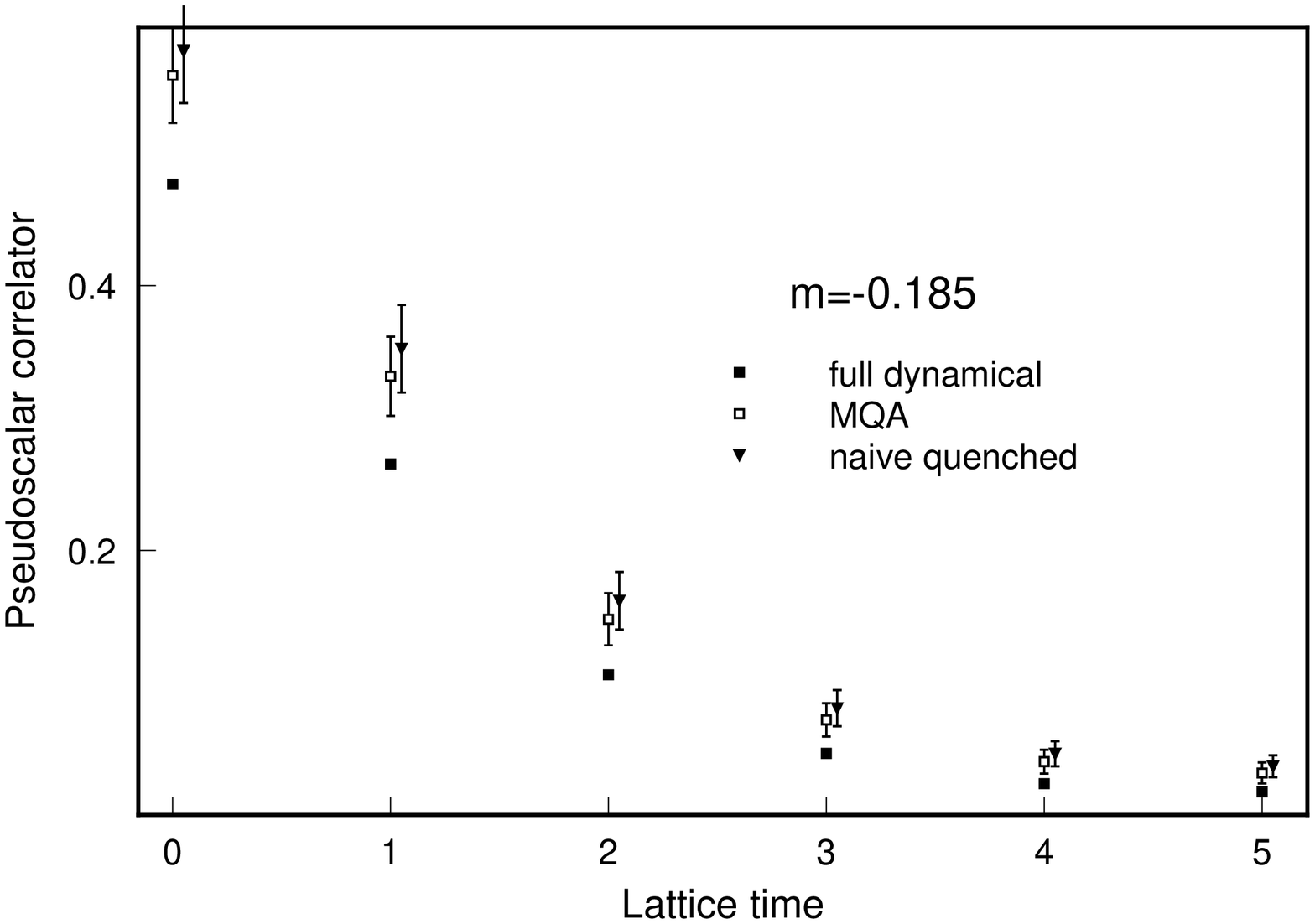,
width=0.75\hsize}$$
\caption{Supercritical pseudoscalar correlators in QED2- quenched, MQA and full dynamical}
\end{figure}

\section{Conclusions}

    The analytic arguments and numerical results presented here provide strong evidence
 for the interpretation of  exceptional configurations as an artifact of quenched Wilson lattice
 theory arising from the structure of the spectrum of the discrete Wilson-Dirac matrix, and
 in particular from the appearance of exact real eigenmodes for particular gauge
 configurations close to the critical line corresponding to the chiral limit. Two
 dimensional QED turns out to be a particularly useful testbed for studying this problem
 for two reasons. First, the zero mode and topological charge structure of the theory
 closely mimics that of 4 dimensional QCD, at least insofar as those aspects germane to
 the real eigenmode problem are concerned. Secondly, extensive simulations can be performed
 at comparatively trivial computational cost (all of the numerical calculations reported in this 
 paper were performed on a Pentium PC), yielding complete information about all aspects
 of the spectral structure. It is also possible to perform full dynamical simulations (with exact
 evaluation of the determinant) in this theory to test the validity of proposed modifications
 of the quenched approximation. 

   Much of the material presented in Sections 2 and 3 extends in a direct and obvious
 fashion to 4 dimensional QCD. There also, discrete symmetries of the Wilson-Dirac
 secular equation  lead to the appearance of exactly real eigenmodes. The quenched
 functional integral as conventionally defined will be  a cut function of the bare quark mass with
 an exceedingly complex branch structure.  Pinch arguments along
 the lines of Section 3 can again be used to 
 identify a subset of branch points.
  Unfortunately, the very useful and detailed 
  information on the lattice spacing,
 lattice volume and clover coefficient dependence of  the real part of the Wilson-Dirac
 spectrum (summarized in the histograms of Figs. 4,5, and 6) is not yet available for
 4 dimensional QCD. However the results presented here for QED2 
 confirm  in all essential details the picture of the zero mode problem in QCD presented 
 in \cite{qzpaper}. In particular, the direct comparisons of meson correlators in the
 conventional quenched, modified quenched (MQA), and exact dynamical simulations
 are a gratifying confirmation of the efficacy of the MQA method in resolving the 
 exceptional configuration problem at finite lattice spacing.

\vspace{1in}
\section{Acknowledgements}
 The work of W.B. and E.E was performed at the Fermi National Accelerator Laboratory,
 which is operated by Universities Research Association, Inc., under contract
 DE-AC02-76CHO3000. The work of A.D. was supported in part by NSF grant 93-22114. 
 The work of H.T. was supported in part by the Department of Energy under 
 grant DE-AS05-89ER 40518. We are grateful for the continued
assistance of G. Hockney.

 \end{document}